\documentclass[preprint,preprintnumbers,amsmath,amssymb]{revtex4}
\usepackage{graphicx}
\usepackage{bm}

\begin{document}

\title{Quantum communication in the spin chain with multiplespin exchange interaction}

\author{Xiang Hao}
\altaffiliation{Corresponding
author,Email:110523007@suda.edu.cn,Phone:86051265226983}

\affiliation{Department of Physics, School of Mathematics and
Physics, Suzhou University of Science and Technology, Suzhou,
Jiangsu 215011, People's Republic of China}

\author{Shiqun Zhu}

\affiliation{School of Physical Science and Technology, Suzhou
University, Suzhou, Jiangsu 215006, People's Republic of China}

\begin{abstract}

The transmission of quantum states in the anisotropic Heisenberg XXZ
chain model with three-spin exchange interaction is studied. The
average fidelity is used to evaluate the state transfer. It is found
out that quantum communication can be enhanced by the anisotropic
coupling and multiple spin interaction. Such spin model can reduce
the time required for the perfect state transmission where the
fidelity is unity. The maximally entangled Bell states can be
generated and separated from the whole quantum systems.



\end{abstract}

\maketitle

\section{Introduction}

Quantum communication is regarded as an important element of quantum
information processing operations \cite{Nielsen00}. In any scenario,
quantum states can be transferred from one location to another
location through a quantum channel. The quantum channel is often
composed of the qubits linearly connected by Heisenberg
interactions. One qubit carries useful quantum information in the
form of a finite state $|\psi\rangle$ while the other qubits are
initialized into a certain state. The whole system evolves such that
another qubit ends up in the state $|\psi\rangle$. This method does
not require the qubit couplings to be switched on and off.
Therefore, it is suitable to implement in solid-state quantum
systems \cite{Divin98,Divin00}. Currently there are many schemes in
low-dimensional quantum spin models
\cite{Bose03,Subrahmanyam04,Christandl04,Li05,Yung06,Fitzsimons06,Zhang06,Zhou06}.
Much more attention was paid to the investigation of the chains with
only nearest-neighboring spin exchange interaction. In condensed
matter physics, multiple spin exchange interactions are often
present in many real quasi-one-dimensional magnets, especially the
oxides of transition metals. Some theoretical models with three-spin
interactions are extensively used for the description of the
magnetic properties of real magnets
\cite{Tsvelik90,Frahm92,Lou04,Zvyagin05}. These models are necessary
because they can provide the possibility to compare experimental
data with exact solutions for one-dimensional models. They often
exhibit some quantum phase transitions \cite{Yang05,Lou06}.
Therefore, it is of great interest to study the effects of multiple
spin exchange interactions on quantum communication.

In this paper, the anisotropic Heisenberg XXZ chain model with
three-spin interaction serves as the quantum channel for the
transfer of arbitrary single-qubit quantum state. In experiments, a
simple three-spin chain model with effective three-body interactions
can be prepared by the optical lattices of equilateral triangles
\cite{Pachos04}. The simplest case of a three-qubit spin chain is
investigated here because the quantum-state transfer over longer
distance can be generated by extending this chain to the spin
network according to the method in \cite{Christandl04}. The exact
dynamics of this model is analytically given in Sec. II. The general
expression of the average fidelity for quantum transmission can also
be obtained when a certain initial state is chosen. Finally, a brief
discussion concludes the paper.

\section{Quantum state transfer}

The quantum channel in this protocol is the anisotropic Heisenberg
three-spin chain with three-body exchange interactions. This
multiple spin interaction corresponds to the coupling between
next-nearest neighbors adjusted by the middle spin \cite{Lou04,
Pachos04}. The theoretical model is extensively investigated in the
field of quantum phase transitions. The Hamiltonian of the chain in
the open boundary condition can be expressed as
\begin{equation}
H=\frac J4 \left[
\sum_{i=1}^{N-1}(\sigma_{i}^x\sigma_{i+1}^x+\sigma_{i}^y\sigma_{i+1}^y+\Delta
\sigma_{i}^z\sigma_{i+1}^z)+\omega(\sigma_{i-1}^x\sigma_{i}^z\sigma_{i+1}^y-\sigma_{i-1}^y\sigma_{i}^z\sigma_{i+1}^x)
\right]
\end{equation}
where $\sigma_i^{\alpha}(\alpha=x,y,z)$ is the spin operator of the
$i$th qubit in the chain. The parameter $J$ is the nearest-neighbor
Heisenberg coupling and $\Delta$ denotes the anisotropy. Here
$\omega$ represents the relative strength of three-body exchange
interaction. For the convenience, the Planck constant $\hbar$ is
assumed to be one and $|1\rangle,|0\rangle$ are the eigenvectors of
$\sigma^z$ with the corresponding eigenvalues $\pm 1$. In the
following discussion, the simplest case of $N=3$ is analytically
studied. The propagator $\hat{U}(t)$ is used to describe the process
of quantum-state transfer,
\begin{equation}
\hat{U}(t)=\sum_{j}\exp(-iE_{j}t)|\varphi_{j}\rangle \langle
\varphi_{j}|.
\end{equation}
Here $E_{j}$ and $|\varphi_{j}\rangle$ include all the eigenvalue of
the Hamiltonian $H$ and the corresponding eigenvector respectively.
Because the Hamiltonian of the system satisfies the symmetric
property of $[H,S^z]=0$ where the symmetric operator $S^z=\frac
12\sum_i \sigma_{i}^z$ is the $z$-component of the total spin, all
of the eigenvalues and the corresponding degenerate eigenvectors can
be written as,
\begin{align}
E_{0}&=J\Delta ;
|\varphi_{0}\rangle=|111\rangle,|\bar{\varphi}_{0}\rangle& \nonumber \\
E_{k=1,2,3}&=\frac {J}{2}\eta_{k} ;
|\varphi_{k}\rangle=a_{k}|100\rangle+b_{k}|010\rangle+c_{k}|001\rangle,|\bar{\varphi}_{k}\rangle&.
\end{align}
In the above equation,$|\bar{\varphi}\rangle$ is the degenerate
eigenvector which is the spin flipped state of $|\varphi\rangle$.
The coefficients satisfy $a_{k}=\sqrt{\frac
{1-b_k^2}{2}}e^{-i\beta_k}$, $c_{k}=a_{k}^{*}$ and $b_{k}=\frac
{\eta_{k}^{2}-\omega^2}{(\eta_{k}^{2}-\omega^2)^2+2(\eta_{k}^{2}+\omega^2)}$
where $\tan \beta_k=\frac {\omega}{\eta_k}$. Here
$\eta_{k}(k=1,2,3)$ belong to the three real roots of the equation
of $\eta^3+2\Delta \eta^2-(2+\omega^2)\eta-2\Delta\omega^2=0$. For
the special case of $\Delta=0$, $\eta_k=0,\pm \sqrt{2+\omega^2}$. In
the product Hilbert space of
$\{|111\rangle,|110\rangle,|101\rangle,|100\rangle,|011\rangle,|010\rangle,|001\rangle,|000\rangle
\}$, the propagator can be written as
\begin{equation}
\hat{U}(t)=\left(\begin{array}{cccccccc}
            e^{-iJ\Delta t}&0&0&0&0&0&0&0\\
            0&\tau_1&\tau_4&0&\tau_6&0&0&0\\
            0&\tau_2&\tau_5&0&\tau_4&0&0&0\\
            0&0&0&\tau_1&0&\tau_2&\tau_3&0\\
            0&\tau_3&\tau_2&0&\tau_1&0&0&0\\
            0&0&0&\tau_4&0&\tau_5&\tau_2&0\\
            0&0&0&\tau_6&0&\tau_4&\tau_1&0\\
            0&0&0&0&0&0&0&e^{-iJ\Delta t}
            \end{array}\right).
\end{equation}
Here the elements are obtained by $\tau_1=\sum_k e^{-iE_k t}a_k
c_k$, $\tau_2=\sum_k e^{-iE_k t}a_k b_k$, $\tau_3=\sum_k e^{-iE_k
t}a_k^2 $, $\tau_4=\sum_k e^{-iE_k t}b_k c_k$, $\tau_5=\sum_k
e^{-iE_k t}b_k^2 $ and $\tau_6=\sum_k e^{-iE_k t}c_k^2$.

In the recent experiments of nuclear magnetic resonance
\cite{Gershenfeld97,Lee07,Du03,Vandersypen04}, the three-body
exchange interactions can be controlled by the selection of
sequences of radio-frequency pulses. By the method in
Ref.\cite{Gershenfeld97,Lee07}, we can initialize the whole system
of three spins $A,B,C$ into the pure state
$|\Psi(0)\rangle=|\psi\rangle_{A}\otimes |00\rangle_{BC}$ where
$|\psi\rangle=\cos\frac {\theta} 2|0\rangle_A+\sin\frac {\theta}2
e^{i\phi}|1\rangle_A,(0\leq\theta\leq \pi,0\leq\phi\leq 2\pi)$ is an
arbitrary single-qubit state. In this scheme, this quantum state
need be transferred from qubit $A$ to qubit $C$. To evaluate the
efficiency of quantum communication, we use the average fidelity,
\begin{equation}
F_{A}(t)=\frac {\displaystyle \int_{0}^{2\pi}d\phi\!\int_{0}^{\pi}
F(t)\,\sin\theta\,d\theta} {4\pi}
\end{equation}
According to Ref. \cite{Nielsen00}, the fidelity for a pure input
state
$F(t)=\{\mathrm{Tr}[\sqrt{(\rho_{0})^{1/2}\rho_{t}(\rho_{0})^{1/2}}]\}^{2}=\mathrm{Tr}[\rho_{t}\rho_{0}]$.
Here $\rho_{0}=|\psi\rangle \langle \psi|$ and
$\rho_{t}=\mathrm{Tr_{AB}}[\hat{U}(t)|\Psi(0)\rangle \langle
\Psi(0)|\hat{U}^{\dag}(t)]$ where the symbol $\mathrm{Tr_{AB}}$
describes the trace over the qubits $A,B$. It is found out that the
average fidelity for the case of
$|\Psi(0)\rangle=|\psi\rangle_{A}\otimes |00\rangle_{BC}$ can be
given by
\begin{equation}
F_{A}(t)=\frac 13\left[ 1+|\tau_6|^2+\mathrm{Re}(e^{iJ\Delta
t}\tau_6) \right]+\frac 16(|\tau_1|^2+|\tau_4|^2).
\end{equation}
Here $\mathrm{Re}$ denotes the real part of a complex number. The
numerical results of the average fidelity can be illustrated in Fig.
1(a) when the anisotroy $\Delta$ is varied. It is clearly shown that
the curve of $F_{A}$ changes periodicly with the time for
$\Delta=0$. Compared to the special case of $\Delta=0$, the
anisotropy $\Delta= 1$ can increase the maximal value of $F_A$. It
is proven that the anisotropy can enhance the efficiency of the
quantum-state transfer. It is interesting to study the quantum
transfer in the case of $N>3$. When the initial state of the system
is $|1\rangle_A \otimes |0\cdots 0\rangle$, the fidelity can be
calculated and shown in Fig. 1(b). The values of the fidelity can be
increased after the long time but the maximal value is always less
than one. The efficiency of the quantum communication is degraded
with the increase of the number of spins.

In order to demonstrate the effects of the multiple spin exchange
interactions on quantum communication, we mainly discuss the special
case of $\Delta=0$. From the expression of the propagator, it is
found out that there is the charateristic time $t_c$ where the
elements satisfy $\tau_1(t_c)=\tau_4(t_c)=0$, $\tau_6(t_c)=-1$ and
$\tau_3(t_c)=\tau_{5}(t_c)$. By the analytical method, we can deduce
the finite expression of charateristic time,
\begin{equation}
t_c=\frac {\displaystyle
(4n+2)\pi+2\arctan{(\omega\sqrt{\omega^2+2})}} {J\sqrt{\omega^2+2}}
,(n=0,1,2,\ldots).
\end{equation}
Owing to the periodic property of the charateristic time, the
average fidelity $F_A$ evolves periodicly like the curve in Fig. 1.
We numerically calculate the minimum of the charateristic time
$t_c(n=0)$ with the variation of the multiple spin exchange
interaction $\omega$ in Fig. 2. It is clearly seen that the values
of $t_c(n=0)$ slightly increase and then decrease with increasing
the relative strength of the multiple spin exchage interaction. When
$\omega=\omega_0\thickapprox 0.6$, the value of the charateristic
time is maximal. When the initial state is chosen as
$|\Psi(0)\rangle=|\psi\rangle_{A}\otimes |00\rangle_{BC}$, the state
at $t_c$ can be expressed as
$|\Psi(t_c)\rangle=|00\rangle_{AB}\otimes
(-\sigma_C^{z}|\psi\rangle_{C})$. Though the average fidelity
$F_{A}(t_c)$ cannot arrive at the maximal value of unity, the
quantum state $|\psi\rangle$ can also be constructed after we apply
the local operation $\sigma_{C}^{z}$ on the qubit $C$. At the
characteristic time, the perfect quantum-state transmission of
$F_A=1$ can also be implemented with no local quantum operations.
For example, if the whole system is initialized into
$|\Psi(0)\rangle=|\psi\rangle_{A}\otimes |01\rangle_{BC}$, the total
state at the charateristic time $t_c$ can be obtained by
\begin{equation}
|\Psi(t_c)\rangle=\frac
{1}{\omega^2+1}\left[(\omega^2-1)|10\rangle_{AB}+2i\omega|01\rangle_{AB}
\right]\otimes |\psi\rangle_{C}
\end{equation}
In this condition, the time required for the perfect quantum
communication can be shortened by increasing the multiple spin
exchange interaction $\omega > \omega_0$. Meanwhile, it is seen that
qubits $A$ and $B$ are entangled and separated from qubit $C$ at the
charateristic time $t_c$. According to Ref.
\cite{Wang02,Bayat05,Wang01}, the concurrence $C$ can be used to
measure the entanglement of the pure state in qubits $A$ and $B$,
\begin{equation}
C=2\max \{0, \frac {\displaystyle
{2\omega(\omega^2-1)}}{(\omega^2+1)^2} \}
\end{equation}
The influence of the multiple spin exchange interactions $\omega$ on
the entanglement $C$ is clearly shown in Fig. 3. The values of $C$
can arrive at the maximal value of one when $\omega\thickapprox
0.42$ or $2.41$. In this condition, the pure state of qubits $A$ and
$B$ is the maximally entangled Bell state which is very useful in
other tasks of quantum information processing. When the values of
$\omega$ are chosen as $0$ and $1$, the pure state corresponds to
the separate state $|10\rangle_{AB}$ and $|01\rangle_{AB}$
respectively. The entanglement will be extinguished if the values
$\omega \rightarrow \infty$. In association with the property of the
charateristic time, it is found out that the time for the
achievement of perfect quantum communication is very short in the
condition of $\omega\thickapprox 2.41$ where the maximally entangled
state can be generated and separated from the system.

\section{Discussion}
An arbitrary single-qubit quantum state is transferred through the
quantum channel of anisotropic Heisenberg three-spin chain with
multiple spin exchangle interaction. The propagator for the
quantum-state transfer is analytically deduced. It is found out that
the anisotropy can improve the efficiency of quantum communication.
For the sepecial case of $\Delta=0$, there exists the charateristic
time where the perfect quantum communication can be carried out.
Under the influence of the multiple spin exchange interaction, the
charateristic time can be reduced by the increase of $\omega$. The
entangled pure state is generated and separated from the quantum
system. This resource is usefully implemented in other tasks of
quantum information processing.

\section{Acknowledgements}

The work was supported by the Research Program of Natural Science
for Colleges and Universities in Jiangsu Province Grant No.
09KJB140009 and the National Natural Science Foundation Grant No.
10904104.

\newpage

{\Large Figure caption}

Figure 1

The solid lines denote the case of the anisotropy $\Delta=0$ and the
dashed lines represent the case of $\Delta=1$. (a): The average
fidelity $F_A$ is plotted as a function of time $t$ when
$J=1,\omega=1$; (b): The fidelity $F$ in the chain with $N=6$ is
calculated when $J=1,\omega=1$.

Figure 2

The minimal value of the charateristic time $t_c(n=0)$ is plotted
with the variation of the multipe spin exchange interaction $\omega$
for the case of $J=1$.

Figure 3

The entanglement $C$ at the charateristic time $t_c$ is plotted as a
function of the multiple spin exchange interaction $\omega$ if the
initial state is chosen as $|\Psi(0)\rangle=|\psi\rangle_{A}\otimes
|01\rangle_{BC}$.

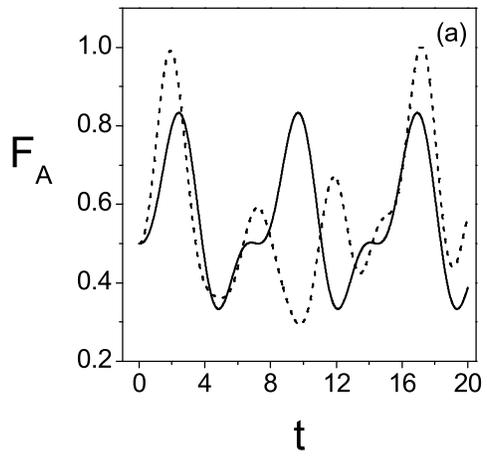
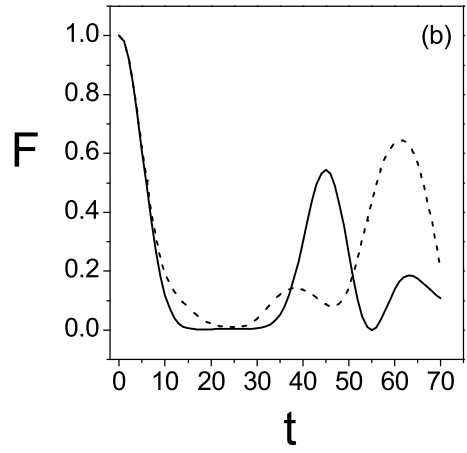
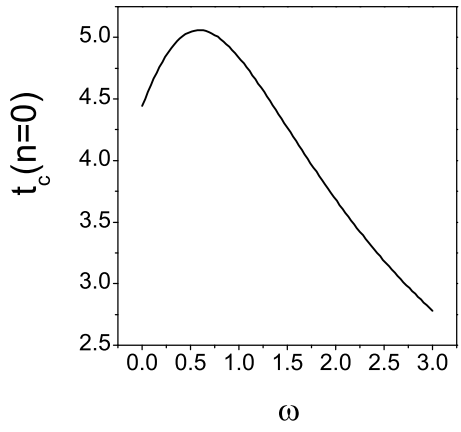
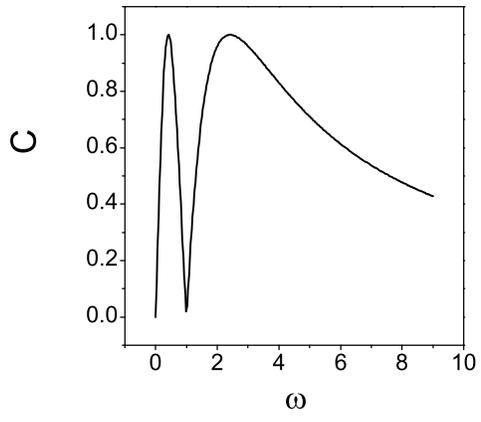
\end{document}